\documentclass[twocolumn,showpacs,preprintnumbers,amsmath,amssymb]{revtex4}

\usepackage{graphicx}
\usepackage{dcolumn}
\usepackage{bm}

\begin{document}
\bibliographystyle{naturemag}


\title{Measuring the role of surface chemistry in silicon microphotonics}
\author{Matthew Borselli}
\email{borselli@caltech.edu}
\author{Thomas J. Johnson}
\author{Oskar Painter}
\affiliation{Department of Applied Physics, California Institute of Technology, Pasadena, CA 91125, USA.}
\homepage{http://copilot.caltech.edu}
\date{\today}

\maketitle

\textbf{\small{The silicon/silicon dioxide (Si/SiO$_{2}$) interface plays a crucial role in the performance, cost, and reliability of most modern
    microelectronic devices\cite{Yablonovitch1,Chabal,Fenner,Yamashita,Kobayashi,Linnros,Li,Petitdidier}, from the basic transistor to flash memory,
    digital cameras, and solar cells.  Today the gate oxide thickness of modern transistors is roughly 5 atomic layers, with 8 metal wire
    layers required to transport all the signals within a microprocessor.  In addition to the increasing latency of such reduced-dimension metal
    wires, further ``Moore's Law'' scaling of transistor cost and density is predicted to saturate in the next decade\cite{Muller2}.  As a result,
    silicon-based microphotonics is being explored for the routing and generation of high-bandwidth
    signals\cite{Liu,Xu1,Boyraz,Rong2,Notomi1,Notomi3,Song}.  In comparison to the extensive knowledge of the electronic properties of the Si/SiO$_2$
    interface, little is known about the optical properties of Si surfaces used in microphotonics.  In this Letter, we explore the optical properties
    of the Si surface in the telecommunication-relevant wavelength band of $\lambda=1400$-$1600$ nm.  Utilizing a high quality factor
    ($Q\sim1.5\times10^6$) optical microresonator\cite{Vahala1} to provide sensitivity down to a fractional surface optical loss of
    $\alpha'_{s}\sim10^{-7}$, we show that optical loss within Si microphotonic components can be dramatically altered by Si surface preparation, with
    $\alpha'_{s}\sim 2{\times}10^{-5}$ measured for chemical oxide surfaces as compared to $\alpha'_{s}\le 2{\times}10^{-6}$ for hydrogen-terminated
    Si surfaces.  These results indicate that the optical properties of Si surfaces can be significantly and reversibly altered by standard
    microelectronics treatments, and that stable, high optical quality surface passivation layers will be critical in future Si micro- and
    nano-photonic systems.}}

\begin{figure}
  \includegraphics{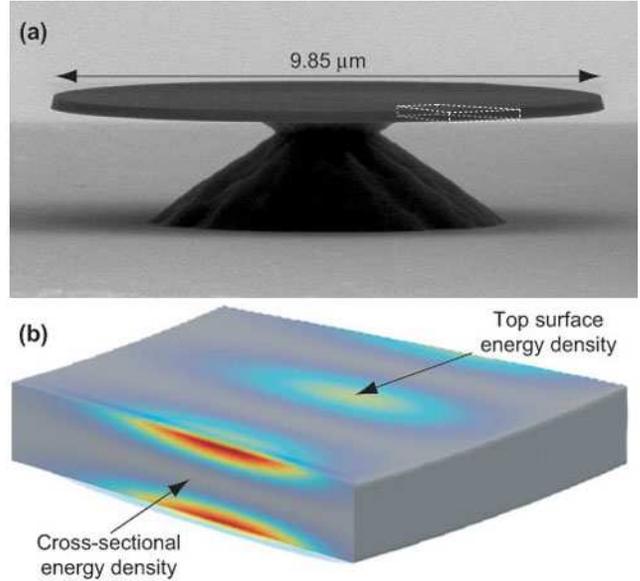}
\caption{\label{fig:mode_composite} Composite of surface-sensitive thin-disk optical resonance. (a) SEM micrograph of a 5 $\mu$m radius SOI
  microdisk. (b) Zoomed-in representation of disk edge (white dashed box) showing a TM polarized whispering gallery mode solved via FEM, indicating large
  electric field energy density at the top and bottom surface of the silicon active layer.}
\end{figure}

Historically, studies of Si surface and interface states have primarily focused on their electronic properties, with sensitive techniques such as
deep-level transient spectroscopy\cite{Lang1} or surface-sensitive minority carrier lifetime measurements\cite{Linnros} being employed.  Three
exceptions to this are deep-level optical spectroscopy\cite{Chantre1}, cavity-ringdown spectroscopy\cite{Aarts}, and the ultra-sensitive
technique of photothermal deflection spectroscopy (PDS)\cite{WB_Jackson1,Amato1} which can measure fractional optical absorption down to $\alpha l
\sim 10^{-8}$.  None of the aforementioned techniques, however, is well suited for studying as-processed microphotonic elements.  In this work we
utilize a specially designed microdisk optical resonator to study the optical properties of surfaces typical in silicon-on-insulator (SOI)
microphotonic elements in a noninvasive, rapid, and sensitive manner.  Shown in Figure \ref{fig:mode_composite}, the high quality factor ($Q$) Si
microdisk resonators used in this work provide surface-specific optical sensitivity due to the strong overlap of the top and bottom surfaces of the
active Si layer with the electric field energy density of appropriately polarized bound optical modes of the microdisk.  In addition, light within
these micron-scale structures circulates tens of thousands of times, providing an effective path length approaching one meter.

\begin{figure*}[ht]
  \includegraphics{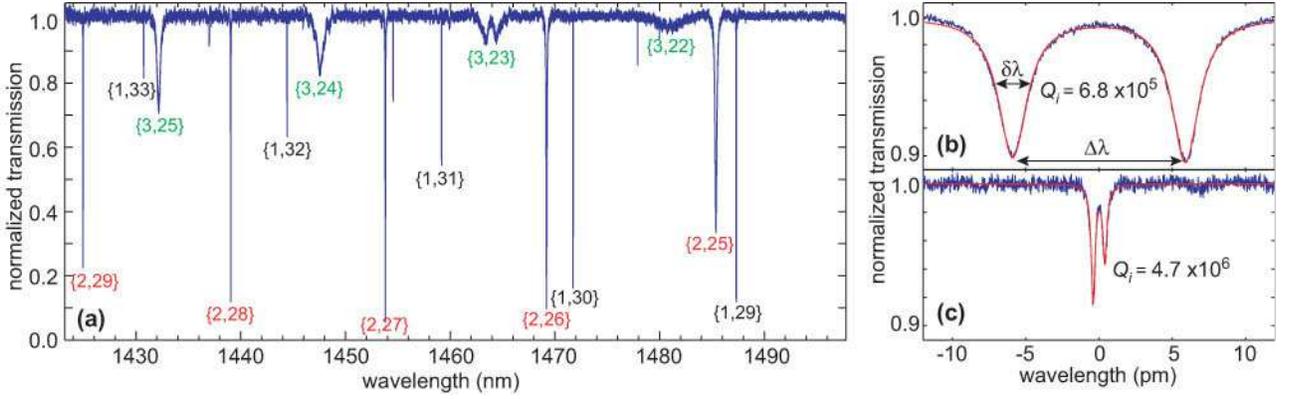}
\caption{\label{fig:full_scan}Normalized spectral transmission response of Si microdisk resonators. (a) Broad scan across $\lambda=1400$ nm band for a $5$ $\mu$m radius microdisk with the fiber taper placed
  $0.6\pm0.1$ $\mu$m away from the disk edge and optimized for TM coupling.  The spectrum was normalized to the response of the fiber taper moved $3$
  $\mu$m laterally away from the disk edge.  Each mode is labelled as \{$p,m$\} where $p$ and $m$ are the radial and azimuthal number, respectively.
  (b) High-resolution scan of the TM$_{1,31}$ mode at $\lambda=1459$ nm in (a). $\Delta\lambda$ and $\delta\lambda$ indicate the CW/CCW mode splitting
  and individual mode linewidth, respectively.  (c) High-resolution scan of a $40$ $\mu$m radius microdisk, showing the reduced loss of a bulk TE WGM.}
\end{figure*}

A normalized measure of surface sensitivity for a guided-wave mode in a waveguide or resonator can be defined as $\Gamma'_s \equiv \Gamma_s / t_s$,
where $\Gamma_s$ is the fractional electric field energy overlap with a surface perturbation of physical depth $t_s$.  If optical loss is dominated by
interactions with the surface, then the modal loss coefficient per unit length ($\alpha_{m}$) measured from experiment can be related to a fractional
loss per pass through the surface given by $\alpha'_{s} = \alpha_{m}/\Gamma'_s$.  As discussed in Ref. \cite{Blood1}, for a true two-dimensional
surface in which the perturbation depth is infinitesimal, $\alpha'_{s}$ is the most relevant quantity describing the surface and is equivalent to the
fraction of power lost for a normal incident plane wave propagating across the surface.  $\alpha'_{s}$ is an important property of all surfaces in
optics, yet it has historically rarely been measured\cite{Amato1}, and is unknown today even for many important surfaces such as the $c$-Si surfaces
studied here.  From finite-element method (FEM) simulations\cite{Spillane3}, shown in Figure \ref{fig:mode_composite}, the transverse magnetic (TM)
polarization whispering-gallery-modes (WGMs) of the microdisk are $\sim90\times$ more sensitive to the top and bottom $\langle100\rangle$ Si surfaces
than the etched sidewall at the microdisk periphery; specifically, $\Gamma'_{{top}}=\Gamma'_{{bot}}=3.5\times10^{-3}$ nm$^{-1}$ and
$\Gamma'_{{side}}=8.1\times10^{-5}$ nm$^{-1}$.  This implies that $\sim0.2\%$ of the optical mode exists in a single monolayer at the top (bottom) Si
surface, while little of the mode sees imperfections at the microdisk perimeter.  At this level of sensitivity, a WGM with a $Q=1.5\times10^6$
(similar to the values measured in devices described below) can be used to measure fractional loss through the Si surface as small as $\alpha'_{s}
\sim 10^{-7}$.  Such high-$Q$ resonators can also be used to measure dispersive effects\cite{Teraoka} of the Si surface chemistry with a sensitivity
corresponding to $0.04\%$ of a Si monolayer, or roughly $\sim10^4$ Si atoms for a $5$ $\mu$m radius microdisk.

The silicon microdisks studied in this work were fabricated from an SOI wafer commercially available from SOITEC, consisting of a $217$ nm thick
silicon device layer ($p$-type, $14-20~\Omega\cdot$cm resistivity, $\langle100\rangle$ orientation) with a $2~\mu$m SiO$_{2}$ buried oxide (BOX)
layer.  Microdisks of $5$ and $10$ $\mu$m radius were fabricated\cite{Borselli2}, finishing with a $10$ minute acetone soak and Piranha etch to remove
organic materials.  A $1$ hour dilute hydrofluoric acid (HF) solution comprised of five parts $18.3$ M$\Omega$ deionized (DI) water to one part
concentrated aqueous HF ($49$\%) was used to remove a protective SiN$_{x}$ cap and partially undercut the disk, as shown in the SEM micrograph in
Figure \ref{fig:mode_composite}(a).  The wafer was then rinsed in deionized water, dried with nitrogen (N$_2$), and immediately transferred into an
N$_2$ purged testing enclosure.

The microdisk resonators were characterized using a tunable external-cavity laser ($\lambda=1420$-$1498$ nm, linewidth $<5$ MHz) connected to a
computer-controlled fiber taper waveguide probe\cite{Borselli1}.  The micron-scale fiber taper probe was formed from a standard single-mode optical
fiber and was used to evanescently excite the WGMs of the microdisk with controllable loading.  Figure \ref{fig:full_scan} shows the normalized
spectral transmission response of a 5 $\mu$m radius microdisk resonator, illustrating clear families of modes having similar linewidth,
$\delta\lambda$, and free-spectral-range (FSR).  By comparison to FEM simulations of the Si microdisk, each mode in Figure \ref{fig:full_scan} was
categorized and labelled as TM$_{p,m}$, where $p$ and $m$ are the radial and azimuthal number, respectively.

Owing to their large surface sensitivity (see Figure \ref{fig:mode_composite}), the spectral signature of the TM$_{1,m}$ modes was used to determine
the quality of the optical surfaces.  Figure \ref{fig:full_scan}(b) shows a high resolution scan across the TM$_{1,31}$ mode.  The observed double
resonance dip, termed a doublet, is a result of surface roughness coupling of the normally degenerate clockwise (CW) and counter-clockwise (CCW)
propagating WGMs\cite{Gorodetsky, Borselli2, Little2}.  The rate at which photons are back-scattered is quantified by the doublet splitting,
$\Delta\lambda$, while the rate at which photons are lost from the resonator is quantified by the intrinsic linewidth, $\delta\lambda$, of the
individual doublet modes.  From a fit to the transmission spectrum of Fig.  \ref{fig:full_scan}(b), $\Delta\lambda=11.9$ pm and $\delta\lambda=2.2$
pm, corresponding to an intrinsic modal quality factor of $Q_i\equiv\lambda_0/\delta\lambda=6.8{\times}10^5$ for this TM$_{1,31}$ mode.  This should
be contrasted with the transmission spectrum shown in Fig. \ref{fig:full_scan}(c) for a more confined, and less surface sensitive, TE WGM of a much
larger $40$ $\mu$m radius microdisk.  From the fit parameters ($\Delta\lambda=0.8$ pm, $\delta\lambda=0.3$ pm), the $Q$ of the buried TE mode is
$Q_i=4.7{\times}10^6$, corresponding to a loss per unit length of $\alpha_i=0.13$ dB/cm.  This is nearly an order of magnitude smaller optical loss
than that of the as-processed TM$_{1,m}$ modes, and provides an upper bound on the bulk Si optical loss of the SOI material.

\begin{figure}
\includegraphics{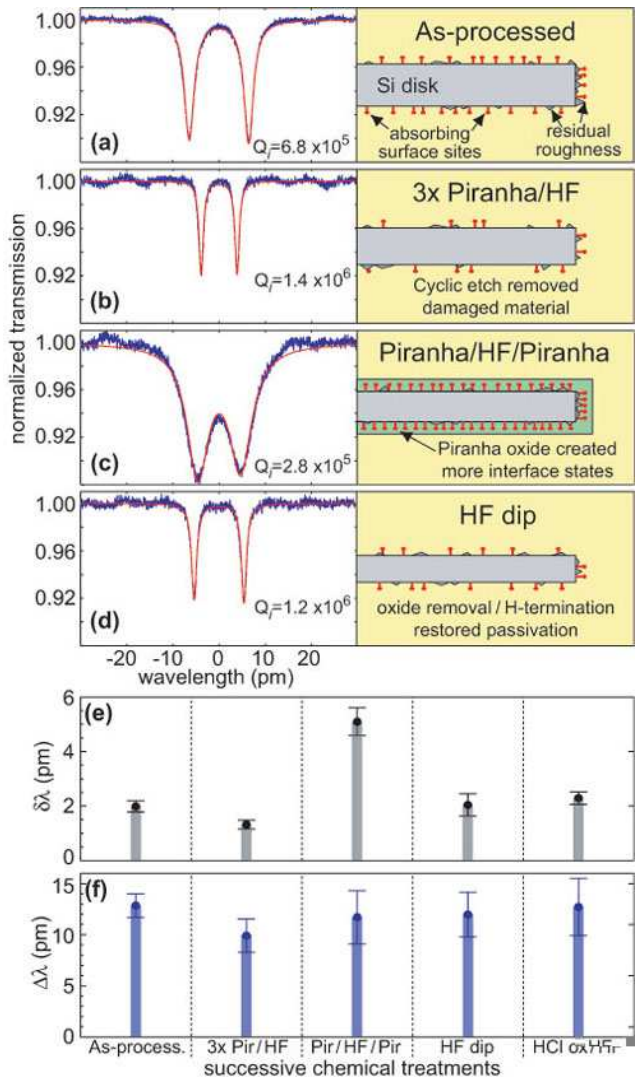}
\caption{\label{fig:chem_treat} (a-d) Taper transmission versus wavelength showing TM$_{1,31}$
  doublet mode after each chemical treatment and accompanying schematic of chemical treatment. (a) As-processed, (b) Triple Piranha/HF cycle
  described in Table \ref{tab:piranha_recipe}, (c) Single Piranha/HF cycle followed by an additional Piranha step allowing controlled measurement of
  piranha oxide, and (d) HF dip to remove chemical oxide from previous treatment and restore passivation. (e) Average intrinsic linewidth,
  $\delta\lambda$ and (f) average doublet splitting, $\Delta\lambda$, for all TM$_{1,m}$ modes within the $1420$-$1470$ nm spectrum after each
  chemical treatment step.}
\end{figure}

The stark difference between the surface-sensitive TM and bulk TE modes indicates that the as-processed Si surfaces are far from optimal.
Etch-induced surface damage on the microdisk sidewall can only account for a small fraction of this difference due to the enhanced sensitivity of the
TM$_{1,m}$ to the top and bottom Si surfaces (comparison of the TM and TE modes in the same microdisk and with similar modal overlap with the
microdisk edge bear this out).  Damage to the top and bottom Si surfaces can stem from a variety of possible sources including chemical mechanical
polishing, native oxide formation during storage, or adventitious organic matter\cite{Fenner}.  In an attempt to repair the Si surfaces a series of
chemical oxidation treatments were performed on the devices.  The well-known process \cite{Chabal,Linnros,Sparacin} of repeated chemical oxidation in
Piranha (H$_2$SO$_4$/H$_2$O$_2$) and HF oxide stripping was employed to controllably prepare the Si surfaces.  Three cycles of the Piranha/HF process,
recipe shown in Table \ref{tab:piranha_recipe}, were applied to the as-processed devices.  From the blue-shift in the WGM resonances due to the three
cycles of the Piranha/HF process, an estimated $1.9\pm0.1$ nm of Si was removed from the surface of the microdisk.  The fit to the TM$_{1,31}$
transmission spectrum, shown in Figure \ref{fig:chem_treat}(b), indicates that a significant improvement to the surfaces has also taken place,
yielding a $\Delta\lambda=7.2$ pm and $\delta\lambda=1.1$ pm.

To separate the effects of the Piranha oxidation and the HF etch, the sample was put through a Piranha/HF/Piranha treatment.  The first cycle of
Piranha/HF was used to ``refresh'' the hydrogen passivation before re-oxidizing the Si surface with Piranha.  Figure \ref{fig:chem_treat}(c) shows the
fit to the now barely resolvable TM$_{1,31}$ doublet yielding $\Delta\lambda=8.6$ pm and $\delta\lambda=5.6$ pm.  The five-fold increase in linewidth
and a negligible increase in doublet splitting is indicative of a significant activation of absorbing surface states without an increase in surface
scattering.  Removing the chemical oxide with the HF dip listed in Table \ref{tab:piranha_recipe} and retesting indicated that an oxide film
equivalent to $2.8\pm0.1$ nm of SiO$_2$ had been present.  The fit to the transmission spectrum of the TM$_{1,31}$ mode in Figure
\ref{fig:chem_treat}(d) yielded fit parameters $\Delta\lambda=9.7$ pm and $\delta\lambda=1.2$ pm, showing that the optical damage to the Si surfaces
caused by Piranha oxidation was reversible.  

As a final treatment to the $5$ $\mu$m radii microdisks, we used the same $3\times$ oxidation and stripping process as described in Table
\ref{tab:piranha_recipe}, but with an HCl based chemistry (8:1:2 H$_2$0:HCl:H$_2$O$_2$, heated to $60^\circ$ C) instead of the H$_2$SO$_4$ based
chemistry.  Figure \ref{fig:chem_treat}(e-f) shows a graphical representation of the average behavior of all TM$_{1,m}$ modes in the $1420$-$1470$ nm
span after each chemical treatment.  The results reveal that the HCl oxidation was slightly less effective at passivating the silicon surface than the
Piranha oxidation; however, it is expected that the optimum solution for chemical oxidation will depend upon the Si crystal orientation and previous
chemical treatments\cite{Petitdidier,Kobayashi}.

\begin{table}
\caption{\label{tab:piranha_recipe}Summary of successive steps for a piranha oxidation surface treatment}
\begin{tabular}{lcrr}
Step & Composition\footnote{All ratios are quoted by volume of standard concentration aqueous solutions} & Temp. & Time\\
\hline
Piranha & 3:1 H$_2$SO$_4$/H$_2$O$_2$ & $100^{\circ}$C & 10 min\\
Rinse & DI H$_2$O & $23^{\circ}$C & 30 sec\\
Rinse & DI H$_2$O & $23^{\circ}$C & 30 sec\\
Rinse & DI H$_2$O & $23^{\circ}$C & 30 sec\\
HF dip & 10:1 H$_2$O/HF & $23^{\circ}$C &  1 min\\
Rinse & DI H$_2$O & $23^{\circ}$C & 15 sec\\
Rinse & DI H$_2$O & $23^{\circ}$C & 15 sec\\
\end{tabular}
\end{table}

Although it has recently been observed\cite{Sparacin} that repeated chemical oxidation and removal of silicon can provide a smoothing action on etched
sidewalls, the large shifts in optical loss with chemical treatment described above can be linked to surface-state absorption as opposed to
surface-scattering.  Whereas the highly confined Si waveguide measurements to date have been sensitive to changes in loss as low as $1$ dB/cm, the
microdisks of this work are sensitive to changes of loss more than an order of magnitude smaller ($<0.03$ dB/cm) where surface chemistry is more likely
to play a role.  Indeed, as mentioned above the TM-polarized microdisk WGMs are selectively sensitive to the top and bottom Si surfaces which are
extremely smooth in comparison to etched surfaces.  The negligible change in average mode-splitting, $\Delta\lambda$, with chemical treatment (Figure
\ref{fig:chem_treat}(f)) is also indicative of little change in surface roughness.

\begin{figure}
\includegraphics{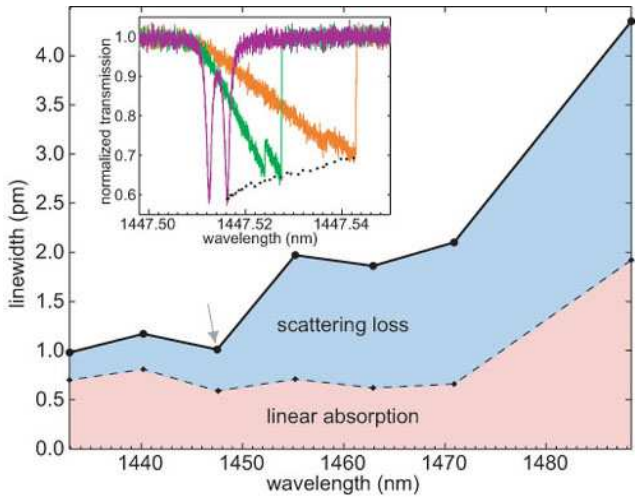}
\caption{\label{fig:wave_depend_loss} Wavelength dependent intrinsic linewidth for a $10$ $\mu$m radius microdisk.  Also shown is the delineation
  between scattering loss versus linear absorption after six cycles of the Piranha/HF treatment given in Table
  \ref{tab:piranha_recipe}.  (inset) Examples of the power dependent transmission versus wavelength data used to
  separate the absorption from the total loss for the TM$_{1,m}$ mode at $\lambda=1447.5$ nm.}
\end{figure}

In order to confirm the above assumption, and to measure the efficacy of the surface-state passivation of the Piranha/HF chemistry, the wavelength
dependent optical quality was measured on $10$ $\mu$m radii disks subjected to a $6\times$ Piranha/HF treatment.  FEM simulations show that $10$
$\mu$m radii disk have approximately the same fractional energy overlap with the top/bottom Si surfaces
($\Gamma'_{{top}}=\Gamma'_{{bot}}=3.7\times10^{-3}$ nm$^{-1}$) as $5$ $\mu$m radii disks but have roughly half the fractional energy overlap with the
etched sidewalls ($\Gamma'_{{side}}=3.8\times10^{-5}$ nm$^{-1}$).  Figure \ref{fig:wave_depend_loss} shows a plot of the wavelength dependent loss
across the full $1420-1500$ nm span for the TM$_{1,m}$ modes of a $10$ $\mu$m radius disk, showing a strong trend towards increasing optical loss with
wavelength.  The black line indicates the intrinsic linewidth for each $m$ number, with an error bar of $\pm0.05$ pm per point (not shown).  The
separation of scattering loss and linear absorption was obtained using a modified version of the power dependent measurements described in Reference
\cite{Borselli2}.  This technique uses the local temperature increase of the silicon microdisk to determine the absorption component of loss, similar
to PDS, but without the need for a thermal model for heat flow; rather, the onset of two-photon absorption changes the fraction of absorption as a
function of power, allowing the linear absorption coefficient to be ascertained directly.  The inset to Fig.  \ref{fig:wave_depend_loss} shows three
examples of power dependent transmission spectra for the $1447.5$ nm mode (grey arrow), along with corresponding transmission minima versus wavelength
(black dots).

From the optical loss measurements of Fig. \ref{fig:wave_depend_loss} it is clear that a significant fraction of residual optical loss, after
Piranha/HF treatment and hydrogen surface passivation, is still due to surface-state absorption (bulk absorption is negligible at this level).  The
strong trend of increasing optical loss with wavelength is seen to be dominated by elastic scattering and/or radiation loss, with the surface-state
absorption being relatively constant across the $\lambda=1420$-$1500$ nm spectrum.  Through measurements of the radiation limited quality factors of
TM$_{1,m}$ modes in the longer wavelength $1565$-$1625$ nm band, intrinsic radiation losses from the circular microdisk are found to contribute
negligibly to the losses in Fig. \ref{fig:wave_depend_loss}.  FEM simulations also show that $\Gamma'$ on all of the surfaces changed less than $5\%$
over the $\lambda=1420$-$1500$ nm span.  The most likely explanation for the observed wavelength dependant loss is improved phase-matching of
surface-scattering into radiation modes with increasing wavelength.

By comparing the cavity $Q$ before and after the Piranha oxide removal, a fractional surface absorption loss per pass of $\alpha_{s,{ox}}^{'}\sim
2{\times}10^{-5}$ is estimated for the Piranha oxide.  This large fractional absorption in the $\lambda=1400$ nm wavelength band ($\hbar\omega\sim
0.85$ eV) is attributed to single-photon absorption by mid-gap interface states.  Such electronic interface states at the Si/(Piranha)SiO$_x$
interface have been observed in Ref. \cite{Yamashita}, with three sets of state-density maxima in the bandgap of silicon occurring at $0.3$, $0.5$,
and $0.7$ eV above the valence-band maximum.  In comparison, from Figure \ref{fig:wave_depend_loss}, the modal absorption loss of the
hydrogen-passivated Si surface is as small as $\alpha_{m}^{H} \sim 0.08$ cm$^{-1}$, corresponding to a fractional surface absorption loss per pass of
$\alpha_{s,H}^{'} \sim 2{\times}10^{-6}$ for the top (bottom) Si active layer surface.

All of the measurements described above were performed in a N$_2$ purged environment over several weeks.  Even in such an environment, however,
changes in the hydrogen passivated surfaces were observed over times as short as a few days.  Left in an unprotected air environment, degradation of
the optical surface quality was evident in a matter of hours.  Research and development of stable surface passivation techniques optimized for optical
quality, akin to the gate oxides of CMOS microelectronics, will be a key ingredient in the future development of Si photonics.  Our data suggests that
surface chemistry as much as surface roughness will ultimately limit the performance of Si microphotonic devices, and further development of Si
passivation techniques should be able to reduce optical losses by as much as an order of magnitude (towards the bulk $c$-Si limit) while
simultaneously improving the stability and manufacturability of future Si photonic components.

\bibliography{sid_v5}
\end{document}